\documentstyle[aps,multicol]{revtex}
\input{epsf}

\newcommand{\be}{\begin{equation}}
\newcommand{\ee}{\end{equation}}
\newcommand{\bea}{\begin{eqnarray}}
\newcommand{\eea}{\end{eqnarray}}
\newcommand{\nn}{\nonumber}

\newcommand{\la}{\langle}
\newcommand{\ra}{\rangle}

\newcommand{\lb}{\left(}
\newcommand{\rb}{\right)}

\newcommand{\mO}{{\mathcal O}}

\begin{document}
\draft
\title{Relative Entropies in Conformal Field Theory}
\author{Nima Lashkari}
\address{
Stanford Institute for Theoretical Physics and Department of Physics,\\ Stanford University, Stanford, CA 94305-4060, USA\\
\smallskip
 Department of Physics and Astronomy, University of British Columbia,\\
6224 Agricultural Road,
Vancouver, B.C., V6T 1W9, Canada\\
}
\maketitle
\begin{abstract}
Relative entropy is a measure of distinguishability for quantum states, and plays a central role in quantum information theory. The family of Renyi entropies generalizes to Renyi relative entropies that include as special cases most entropy measures used in quantum information theory. We construct a Euclidean path-integral approach to Renyi relative  entropies in conformal field theory, then compute the fidelity and the relative entropy of states in one spatial dimension at zero and finite temperature using a replica trick. In contrast to the entanglement entropy, the relative entropy is free of ultraviolet divergences, and is obtained as a limit of certain correlation functions. The relative entropy of two states provides an upper bound on their trace distance. \end{abstract}
 
\begin{multicols}{2}

In the past few years we have witnessed a rapid growth in applications of information-theoretic techniques to field theory, and condensed matter. Most of these applications focus on studying the entanglement entropy, defined as the von Neumann entropy of the reduced density matrix on spatial regions. The entanglement entropy in the ground state, and low-energy excited states has proven to contain a great deal of information about the universal properties of critical systems, their phase-structure and low-energy dynamics; see \cite{Calabrese:2009qy} for a review. 

In relativistic field theories the entanglement entropy suffers from ultraviolet divergences due to the entanglement of arbitrarily high energy modes. The divergent part of the entanglement entropy is the same for all finite energy states, and carries no information about the state. This creates a need for other entropy concepts that are divergence-free. In this work we generalize the technique introduced in \cite{Holzhey:1994we,Calabrese:2004eu} to compute in field theory a large class of divergence-free entropic measures known as Renyi relative entropies that include as special cases most entropies studied in information theory. Assuming analyticity of Renyi relative entropy we then extract the relative entropy of two density matrices defined as:
\bea
S(\rho\|\sigma)=\text{tr}(\rho\log\rho)-\text{tr}(\rho\log\sigma).
\eea
Relative entropy is a measure of the distinguishability of two states in the asymptotic limit of a large number of copies \cite{Ohya,relentropy}. Special cases of relative entropy include the entanglement entropy, the mutual information, and the conditional entropy.
The ultraviolet divergence in entanglement entropy can be attributed to the fact that the maximally mixed density matrix is not a finite energy state in relativistic field theories. All physically-relevant contributions to the entanglement entropy can be thought of as special cases of relative entropy.\footnote{The finite piece of the entanglement entropy of a state $\rho$ appears in its relative entropy with respect to the Gibbs state of the same energy: $S(\rho)=S_{ther}-S(\rho\|\rho_{ther})$, where $\text{tr}(\rho H)=\text{tr}(\rho_{ther} H)$. 
}

The relative entropy can be expressed as $S(\rho\|\sigma)=\Delta \la H_\sigma\ra-\Delta S$, where $\Delta S$ is the difference in von Neumann entropy of $\rho$ and $\sigma$, and $H_\sigma=-\log\sigma$ is the modular Hamiltonian (entanglement Hamiltonain) of $\sigma$. The relative entropy of states with respect to the Gibbs state, $\sigma_A=e^{-\beta H_A}/Z$, is the standard free energy. In this letter, we evaluate the relative entropy of arbitrary states with respect to restrictions of both vacuum and finite temperature states to spatial subregions. By analogy, the relative entropies we compute here can be thought of as generalizations of free energy.

The quantum Renyi relative entropies of two states $\rho$, and $\sigma$ are defined to be \cite{Mark,frederic}
\bea\label{relentropy}
&&S_\alpha(\rho\|\sigma)=\left\{ 
  \begin{array}{l l}
   \frac{1}{\alpha-1}\log\text{tr}\left(\left(\rho_\alpha(\sigma)\right)^\alpha\right) & \quad \text{if $\rho\not\perp \sigma \wedge (\sigma\gg \rho \vee \alpha<1)$}\\
    \infty & \quad \text{else}
  \end{array}
   \right.\nn\\
  &&\rho_\alpha(\sigma)=\sigma^{\frac{1-\alpha}{2\alpha}}\:\rho\:\sigma^{\frac{1-\alpha}{2\alpha}},
    \eea
for any $\alpha\in(1/2,1)\cup(1,\infty)$.\footnote{For the range $\alpha\in(0,1/2)$ it is more natural to use Petz $\alpha$-entropies defined as $\frac{1}{\alpha-1}tr\lb \rho^\alpha\sigma^{1-\alpha}\rb$ \cite{petz}. The definition in (\ref{relentropy}) sometimes referred to as the {\it sandwiched} Renyi relative entropy is a non-commutative generalization of Petz $\alpha$-entropies.} The notation $\sigma\gg\rho$ denotes that $\sigma$ dominates $\rho$, i.e. the kernel of $\sigma$ is contained in the kernel of $\rho$. Although in this work we use Renyi relative entropies as a calculation trick to find the relative entropy and the fidelity of quantum states, they are physically meaningful quantities on their own right. Renyi relative entropies appear naturally in the study of thermodynamics of small systems, in strong converse theorems for coding tasks, and have an operational interpretation in terms of hypothesis testing, respectively \cite{Mark,stephanie,Mosonyi}.

Important special cases of Renyi relative entropies are $\alpha=2$, and $\alpha=1/2$ that are related to the {\it collision relative entropy} and {\it fidelity}, respectively. Quantum fidelity is a natural generalization of the notion of pure states overlap, $|\la\psi|\phi\ra|$, to mixed states, and has proved to be extremely useful in characterizing quantum phase transitions \cite{zanardi}. 
Note that similar to Renyi entropies, under the assumption of analyticity, we can obtain the relative entropy by taking the limit $\alpha\to 1$ of Renyi relative entropies; i.e. $\lim_{\alpha\to 1^\pm}S_\alpha(\rho\|\sigma)=S(\rho\|\sigma)$. In this work, we use this fact to construct a replica trick for relative entropy in conformal field theories. 

%
\section{Relative entropy replica trick}\label{sec1}
Consider a 1+1-dimensional conformal field theory (CFT) on a cylinder of circumference $L$. The reduced density matrix of a region $A=(u,v)$ is $\sigma=\text{tr}_{\bar{A}}|\Omega\ra\la\Omega|$, where $|\Omega\ra$ is the vacuum state. 
Our starting point is a path-integral representation for fractional powers of $\sigma$. Denote by $x$ the dimensionless parameter $|u-v|/L$. Applying the conformal transformation $z=\sin\left[\pi(\omega-u)/L\right]/\sin\left[\pi(\omega-v)/L\right]$ to the cylinder, the density matrix element $\la\phi^+| \sigma|\phi^-\ra$ becomes proportional to the path-integral over the $z$-plane with boundary conditions $\phi(y,0^\pm)=\phi^\pm$, $y\in(-\infty,0)$. Inserting a resolution of the identity at $\text{arg}(z)=\pi(1\pm2\gamma)$ splits $\sigma$ according to:
\bea
\la\phi^+|\sigma|\phi^-\ra=\int{\mathcal D}\phi{\mathcal D}\phi' \la\phi^+|\sigma^\gamma|\phi\ra\la\phi|\sigma^{1-2\gamma}|\phi'\ra\la\phi'|\sigma^\gamma|\phi^-\ra.\nn
\eea

\begin{figure}[t]
\begin{center}
\leavevmode
\epsfxsize=3in
\epsfbox{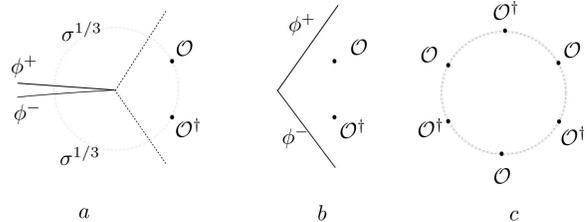}
\end{center}
\caption{($a$) The path-integral for $\rho$ can be written as a product of three matrices: $\sigma^{1/3}\rho_3(\sigma)\sigma^{1/3}$. ($b$) Performing the path-integral only on the slice with operator insertions $\mathcal{O}$ and boundary conditions $\phi^-$ and $\phi^+$ imposed on each end corresponds to computing $\rho_3(\sigma)$.  ($c$) The correlation function found by normalizing $tr\lb\rho_3(\sigma)^3\rb$.}
\label{fig1}
\end{figure}

Now consider the density matrix of an excited state $\rho=tr_{\bar{A}}\lb \mO(-i\infty)|\Omega\ra\la\Omega| \mO^\dagger(i\infty)|\rb$. On the $z$-plane the operators are inserted at $e^{\pm i\pi x}$. For $\alpha<1/x$ the density matrix $\rho$ on the $z$-plane splits as
\bea
&&\la\phi^+|\rho|\phi^-\ra=\nn\\
&&\int{\mathcal D}\phi{\mathcal D}\phi' \la\phi^+|\sigma^{-\frac{1-\alpha}{2\alpha}}|\phi\ra\la\phi|\rho_\alpha(\sigma)|\phi'\ra\la\phi'|\sigma^{-\frac{1-\alpha}{2\alpha}}|\phi^-\ra.\nn
\eea
with the operator insertions contained in $\rho_\alpha(\sigma)$.
From this it is clear that for integer values of $\alpha$, $tr(\rho_n(\sigma)^n)$ is given by a $z$-plane path-integral with $2n$ operator insertions; see figure \ref{fig1}.

In order to compute Renyi relative entropies one has to first properly normalize both $\rho$, and $\sigma$. If $\tilde{\rho}$ are $\tilde{\sigma}$ are the unnormalized states defined by the path-integral then,
\bea\label{rel}
S_n(\rho\|\sigma)&&=\frac{1}{n-1}\log\frac{\text{tr}(\tilde{\rho}_n(\tilde{\sigma})^n)\:(\text{tr}\sigma)^{n-1}}{(\text{tr}\tilde{\rho})^n}\nn\\
&&=\frac{1}{n-1}\log F^\rho_n(\sigma),\nn\\
F_n^\rho(\sigma)&&=\frac{\la \prod_{k=1}^{n}\mO(z_k)\mO^\dagger(z'_k)\ra}{\la \mO(z_0)\mO^\dagger(z'_0)\ra^n}.
\eea
For $\sigma$ the vacuum density matrix and the excited states are primary states we have $z_k=e^{i\pi(\frac{2k+1}{n}+x)}$ and $z'_k=e^{i\pi(\frac{2k+1}{n}-x)}$. The formula in (\ref{rel}) is similar to the expression found for the entropy of excited states in \cite{alcaraz}. However, it is important to notice that the location of operator insertions are not the same.

In the limit $n\to 1$ we obtain the relative entropy of the states $\rho$ and $\sigma$. One might worry that since (\ref{rel}) holds only for $n<1/x$, the limit $n\to 1$ might fail to capture the correct relative entropy. However, as we see explicitly below, this limit reproduces the correct relative entropy for all subsystems of size $0<x<1/2$.

\section{Zero temperature}
As the first example consider free $c=1$ massless bosons on a circle of circumference $L$, and choose the state created by the insertion of a holomorphic vertex operator at infinite past: $V=e^{ia\phi}$, where $\phi$ is the boson field. This is a primary operator of dimension $(h,\bar{h})=(\frac{a^2}{2},0)$. We are interested in computing the relative entropy of the reduced density matrix on a region of size $x L$ in this excited state with respect to the vacuum. Applying a second conformal transformation $\omega=-i\ln z$ maps the $z$-plane to a cylinder of height $2\pi$. Following (\ref{rel}) we would like to compute
\bea
F_n^\rho(\sigma)=\frac{\la \prod_{k=0}^{n-1}V(\pi(\frac{2k+1}{n}+x))V^\dagger(\pi(\frac{2k+1}{n}-x))\ra_{cyl}}{\la V(\pi x)V^\dagger(-\pi x)\ra_{cyl}^n}.
\eea
Correlation functions of vertex operators on a cylinder take the form $\la \prod_kV_{a_k}(t_k)\ra=\prod_{k>i}\left[2\sin(t_{ki}/2)\right]^{-a_k a_i}$. Therefore, one finds $F^\rho_n(\sigma)=\lb n\sin(\pi x)/\sin(n\pi x)\rb^{n a^2}$, and the $n^{th}$ Renyi relative entropy is given by
\bea
S_n(\rho^V_A\|\sigma_A)=\frac{n a^2}{n-1}\log\lb\frac{n\sin(\pi x)}{\sin(n\pi x)}\rb.
\eea 
Analytically continuing in $n$ we obtain the relative entropy, and fidelity of these states:
\bea
&&S(\rho^V_A\|\sigma_A)=a^2(1-\pi x\cot(\pi x)),\nn\\
&&F(\rho^V_A,\sigma_A)\equiv e^{-\frac{1}{2}S_{1/2}(\rho^V_A\|\sigma_A)}=\cos\lb\frac{\pi x}{2} \rb^{a^2/2}.
\eea
In the limit $x\to 1$, the density matrices become orthogonal pure states of the full system, and hence the relative entropy diverges, while the fidelity vanishes as expected. A further consistency check comes from the knowledge of the vacuum modular Hamiltonian: $H_{\Omega}=\frac{L^2}{2}\int_{-x}^{ x}dy \frac{\cos(\pi y)-\cos(\pi x)}{\sin(\pi x)} T_{00}$ \cite{Blanco:2013joa}. The change in the cylinder energy density due to the excitation is $\Delta \la T_{00}\ra=\pi a^2/L^2$. Hence, one finds $\Delta S=\Delta\la H_\Omega\ra-S(\rho^V_A\|\sigma_A)=0$ which matches the result obtained in \cite{alcaraz} using the replica trick for the entanglement entropy of excited states.

Next consider a primary excited state in a generic CFT on a circle: ${\mathcal O}(-i\infty)|\Omega\ra$. We denote the conformal dimension of ${\mathcal O}$ by $(h,\bar{h})$. The calculation of the relative entropy of the reduced density matrix on a subsystem of size $x L$ in this excited state with respect to the vacuum reduces to finding a $2n$-point correlator of ${\mathcal O}$ as a function of $x$. In the limit of small subsystem one can compute $F_n^\rho(x)$ perturbatively in $x\ll 1$ using the operator product expansion: ${\mathcal O}\times {\mathcal O}^\dagger={\bf 1}+{\bf \Psi}+\cdots$. Then, to the first non-trivial order in $x$:
\bea\label{Fn}
F_n^\rho(x)=1+\lb C^\Psi_{\mathcal{O}\mathcal{O}^\dagger}\rb^2 \lb\frac{\prod_{m=1}^{n-1}\sin(\pi m/n)^{n-m}}{(4\pi i x)^{n(n-1)/2}}\rb^{-2(\Delta_\Psi+\bar{\Delta}_\Psi)}\nn
\eea
Notice that the Renyi relative entropies vanish up to the order $O(x^{2(\Delta_\Psi+\bar{\Delta}_\Psi)})$. This can be understood as a consequence of the first law of entanglement thermodynamics  for small subsystems \cite{Bhattacharya:2012mi}. Applying Pinsker's inequality we find that in the limit $x\ll 1$, $\rho_A$ approaches $\sigma_A$ as $\|\rho_A-\sigma_A\|= O(x^{(\Delta_\Psi+\bar{\Delta}_\Psi)})$. In \cite{alcaraz} the change in the entanglement entropy of $A$ due to a primary excitation was computed as a limit of a $2n$-point correlator different from the one in (\ref{Fn}). Here, we prove that both replica tricks produce the same answer. The expression in \cite{alcaraz} for the change in entropy using the entanglement entropy replica trick can be written in terms of $F_n^\rho$ as defined in (\ref{Fn}):
\bea
&&\Delta S=\partial_n\left[2n(h+\bar{h})\log\lb\frac{n\sin(\pi x/n)}{\sin(\pi x)}\rb-\log F_n^\rho(x/n)\right]\Big|_{n\to1}\nn\\
&&=2(h+\bar{h})(1-\pi x\cot(\pi x))-S(\rho_A\|\sigma_A)-x\partial_x\log F_1(x)\nn\\
&&=\Delta \la H_\Omega\ra-S(\rho_A\|\sigma_A),\nn
\eea
where we have used $\lim_{n\to 1}\log F_n(x)=0$, which is a consequence of the finiteness of relative entropy.

\begin{figure}[t]
\begin{center}
\leavevmode
\epsfxsize=3.4in
\epsfbox{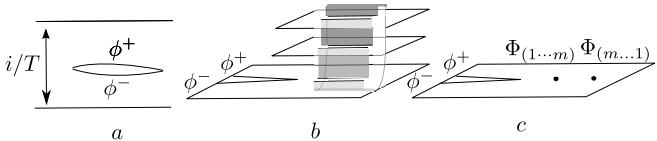}
\end{center}
\caption{($a$) The path-integral corresponding to the reduced density matrix at temperature $T$: $\sigma_T$. ($b$) The density matrix of $\rho_{T/3}$ after the conformal transformation that maps $\sigma_T$ to the complex plane. ($c$) The $m$-sheeted cover can be computed using the correlation function of twist operators.}
\label{fig2}
\end{figure}

\section{Finite temperature}
Up to this point we have only discussed zero temperature states. Now we would like to consider the relative entropy of reduced density matrices on a region $A=(0,l)$ in a CFT on a line at two different temperatures: $T$ and $T_0=T/m$, $m$ an integer. The reduced density matrix of region $A$ in a CFT on a line at finite temperature $T$ is $\sigma_T=\text{tr}_{\bar{A}}\lb e^{-H/T}/Z \rb$. The conformal transformation $z=\sinh(\pi(\omega-l) T)/\sinh(\pi\omega T)$ maps the reduced density matrix at temperature $T$ to the complex plane. 
Under the conformal transformation the identified boundaries at $\omega_{re}\pm i/(2T)$ are mapped to the interval $(e^{-\pi x},e^{\pi x})$ on the real line, where $x=l T$. As we argued, fractional powers of the density matrix, $\sigma_T^\gamma$, are proportional to the path-integral over the region $ 0\leq  \text{arg}(z)\leq 2\pi\gamma$. 

The same transformation acting on the reduced density matrix at temperature $T_0$ maps it to an $m$-sheeted Riemann surface with sheets connected along branch cuts at $(e^{-\pi x},e^{\pi x})$. The partition function on this $m$-sheeted cover is proportional to the two-point correlation function of twist operators inserted at the branch points: $\la \Phi_{(1,2,\cdots,m)}(e^{-\pi x})\Phi_{(m,m-1,\cdots,1)}(e^{\pi x})\ra$; see figure \ref{fig2}. The twist operator $\Phi_{(a_1,a_2,\cdots a_m)}$ sews the sheets according to $a_1\to a_2\to ...\to a_m$.

\begin{figure}[t]
\begin{center}
\leavevmode
\epsfxsize=2in
\epsfbox{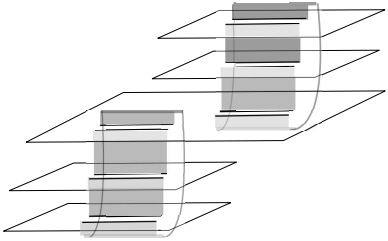}
\end{center}
\caption{The 5-sheeted Riemann surface one constructs to compute the second Renyi relative entropy with $m=3$.}
\label{fig3}
\end{figure}

Following the prescription discussed in section (\ref{sec1}) we cut off fractional powers of the density matrix at temperature $T$ from the path-integral for $\rho_{T_0}$.  Then, gluing $\rho_n(\sigma)$ together we find that $tr(\rho_n(\sigma)^n)$ is a Riemann surface with $q=n(m-1)+1$ sheets: a main sheet with $n$ branch cuts at $e^{i\pi(2k+1)/n}(e^{-\pi x},e^{\pi x})$ for $k=0,\cdots,n-1$, and each cut is glued to $(m-1)$ separate sheets; see figure \ref{fig3}. 

 In the language of twist operators the $n^{th}$ relative entropy is given by 
\bea\label{twist}
&&S_n(\rho_{T/m}\|\sigma_{T})=\frac{1}{n-1}\log\frac{\la \prod_{k=0}^{n-1}\Phi_{(1a^k_2\cdots a^k_m)}(z_k)\Phi_{(a^k_m\cdots a^k_21)}(z'_k)\ra}{\la \Phi_{(1a_2\cdots a_m)}(e^{-\pi x})\Phi_{(a_m\cdots a_21)}(e^{\pi x})\ra^n},\nn\\
&&z_k=e^{\pi\lb i\frac{2k+1}{n}-x\rb},\qquad z'_k=e^{\pi\lb i\frac{2k+1}{n}+x\rb},
\eea
where $a_i^k$ is the $i^{th}$ sheet of the $k^{th}$ branch cut.
All the twist operators in (\ref{twist}) are primaries of weight $\Delta_m=(c/24)(m-1/m)$.
The numerator is a $2n$-point correlation function that is hard to compute in general. From the Riemann-Hurwitz theorem it is clear that our $n(m-1)+1$-sheeted surface is in fact a Riemann sphere. The computation of the correlation function of twist operators in question involves finding the rational conformal map that uniformizes this sphere \cite{Lunin:2000yv}. In the limit  $x=l T\gg 1$ this correlator is dominated by the channel coming from the contractions $\prod_k\Phi_{(1a^k_2\cdots a^k_m)}\sim\Phi_{(1,\cdots ,n(m-1)+1)}$, and $\prod_k\Phi_{(a^k_m\cdots a^k_21)}\sim\Phi_{(n(m-1)+1,\cdots ,1)}$. Our correlator of interest computed in this channel is
\bea
&&\la \Phi_{(1a^1_2\cdots a^1_m)}\Phi_{(a^1_m\cdots a^1_21)}\cdots \Phi_{(1a^n_2\cdots a^n_m)}\Phi_{(a^n_m\cdots a^n_21)}\ra\nn\\
&&\simeq\la \Phi_{(1\cdots n(m-1)+1)}(0)\Phi_{(n(m-1)+1... 1)}(e^{\pi x})\ra\nn\\
&&\simeq c_q e^{-4\pi x\Delta_q},
\eea
for some constant $c_q$ that depends on the OPE coefficients of twist operators.
Similarly at large $x$ the denominator is well approximated by
\bea
\la \Phi_{(1a_2\cdots a_m)}(0)\Phi_{(a_m\cdots a_21)}(e^{\pi x})\ra^n\simeq c_m^n e^{-4\pi x n\Delta_m}.\nn
\eea 

At infinite $x$ the subsystem $A$ has infinite volume. One expects  both $\Delta S$ and $\Delta \la H_\sigma\ra$ to diverge linearly in $x$. 
While there are no ultraviolet divergences in the Renyi relative entropy, there can be an infrared divergent piece due to the infinite volume as in this case:
\bea
S_n(m)\simeq&&\frac{4\pi x}{n-1}\lb n\Delta_m -\Delta_q\rb\nn\\
=&&\frac{\pi c x}{6(n-1)}\lb (n-1)-\frac{n}{m}+\frac{1}{n(m-1)+1}\rb.
\eea
Analytically continuing in $n$, the relative entropy and fidelity are found to be
\bea\label{thermal}
&&S(\rho_{T/m}\|\sigma_T)=\frac{\pi c l T}{6}\lb 1/m-1\rb^2\nn\\
&&F(\rho_{T/m},\sigma_T)=\exp\lb-\frac{\pi c l T}{12}\frac{(1/m-1)^2}{1/m+1}\rb.
\eea
This matches the result of the holographic calculation of the relative entropy on a half-line at finite temperature \cite{Blanco:2013joa}.
The $m\to\infty$ limit corresponds to the vacuum density matrix: $\rho=tr_{\bar{A}}|\Omega\ra\la\Omega|$.

In out-of-equilibrium situations the relative entropy of the reduced density matrix on a region $A$ at time $t$ with respect to $\sigma_T$ puts a lower bound on how far the local state is from equilibrium. This is seen by applying the Pinsker inequality that relates the relative entropy of two states to their trace norm according to: $\|\rho-\sigma\|^2\leq 2S(\rho\|\sigma)$. At late times, the relative entropy is small, and therefore the states are close in trace distance. 

\section{Conclusions}
In this work we have taken a small step towards crossing the language barrier between information theory and field theory by computing a large class of entropic measures in conformal field theories. 
We presented a Euclidean path-integral approach to Renyi relative entropies, and found that the relative entropy of reduced density matrices in excited states with respect to the vacuum or the thermal state are given by certain correlation functions. Note that the relative entropy replica trick exists in all dimensions. However, the correlation functions of interacting theories are hard to compute in higher than one spatial dimension.
 
The entanglement entropy replica trick for disjoint subsystems is discussed in detail in \cite{Faulkner:2013yia,Hartman:2013mia}. The replica trick for the relative entropy of disjoint subsystems provides an interesting Euclidean approach to a direct computation of mutual information, and conditional entropies in terms of the partition function of higher genus Riemann surfaces \cite{Lashkari}. The time-dependence of relative entropy can also be studied using the Euclidean method described here. In systems with dissipation, the relative entropy of $\rho(t)$ with respect to the equilibrium state provides dynamical information on how the system approaches equilibrium.

%
%
%
%

We thank Patrick Hayden and James Sully for discussions.  This work has been supported in part by Natural Sciences and Engineering Research Council of Canada..



\end{multicols}
\end{document}